\documentclass[nofootinbib,pra,superscriptaddress,a4paper,twocolumn]{revtex4-1}
\usepackage{geometry}
\usepackage[english]{babel}
\usepackage[latin1]{inputenc}
\usepackage{hyperref}
\usepackage{amsmath, amssymb, amsfonts, mathrsfs}
\usepackage{graphicx}
\usepackage{multirow}
\usepackage{color, colortbl}
\usepackage{varwidth}
\definecolor{mycolor}{rgb}{0.6, 0.8, 0.2}
\usepackage{capt-of}
\usepackage{diagbox}
\newcommand{\ket}[1]{|#1\rangle}

\newcommand{\braket}[1]{ \langle #1 \, \rangle}

\begin{document}

\title{Phenomenology of complex structured light in turbulent air}

\author{Xuemei Gu}
\email{xmgu@smail.nju.edu.cn}
\affiliation{State Key Laboratory for Novel Software Technology, Nanjing University, 163 Xianlin Avenue, Qixia District, 210023, Nanjing City, China.}
\affiliation{Institute for Quantum Optics and Quantum Information (IQOQI), Austrian Academy of Sciences, Boltzmanngasse 3, 1090 Vienna, Austria.}

\author{Lijun Chen}
\email{chenlj@nju.edu.cn}
\affiliation{State Key Laboratory for Novel Software Technology, Nanjing University, 163 Xianlin Avenue, Qixia District, 210023, Nanjing City, China.}

\author{Mario Krenn}
\email{mario.krenn@univie.ac.at; present address: Department of Chemistry, University of Toronto, Canada \& Vector Institute for Artificial Intelligence, Toronto, Canada.}
\affiliation{Institute for Quantum Optics and Quantum Information (IQOQI), Austrian Academy of Sciences, Boltzmanngasse 3, 1090 Vienna, Austria.}
\affiliation{Vienna Center for Quantum Science \& Technology (VCQ), Faculty of Physics, University of Vienna, Boltzmanngasse 5, 1090 Vienna, Austria.}

\begin{abstract}
	The study of light propagation has been a cornerstone of progress in physics and technology. Recently, advances in control and shaping of light have created significant interest in the propagation of complex structures of light -- particularly under realistic terrestrial conditions. While theoretical understanding of this research question has significantly grown over the last two decades, outdoor-experiments with complex light structures are rare, and comparisons with theory have been nearly lacking. Such situations show a significant gap between theoretical models of atmospheric light behaviour and current experimental effort. Here, in an attempt to reduce this gap, we describe an interesting result of atmospheric models which are feasible for empirical observation. We analyze in detail light propagation in different spatial bases and present results of the theory that the influence of atmospheric turbulence is basis-dependent. Concretely, light propagating as eigenstate in one complete basis is stronger influenced by atmosphere than light propagating in a different, complete basis. We obtain these results by exploiting a family of the continuously adjustable, complete basis of spatial modes -- the Ince-Gauss modes. Our concrete numerical results will hopefully inspire experimental efforts and bring the theoretical and empirical study of complex light patterns in realistic scenarios closer together.
\end{abstract}
\date{\today}
\maketitle

\section{Introduction}
Complex, higher-order spatial modes of light have come under the spotlight due to its inherent orthogonality, and discrete infinite state spaces \cite{yao2011orbital, padgett2017orbital, erhard2018twisted}. Its well-developed technology of generation \cite{beijersbergen1994helical, heckenberg1992generation, vijayakumar2019generation, marrucci2006optical, mirhosseini2013rapid} and manipulation \cite{leach2002measuring, berkhout2010efficient, lavery2012refractive, zhou2017sorting, gu2018gouy} allows the increasing applications of free-space optical (FSO) communication links using spatial modes. The first experiment using spatial modes to encode information in an FSO channel was implemented by Gibson et al. in 2004 \cite{gibson2004free}. Since these early demonstrations, the transmission rate has been increased to 100 Tbit/sec \cite{wang2012terabit, huang2014100}. However, realistic FSO links involve atmospheric turbulence which causes random fluctuations of the intensity and distorts the phase front of the transmitted light beam \cite{andrews2005laser}.

A natural question that arises is, "How does atmospheric turbulence influence complex spatially modulated beams of light?". The answer to this question would not only be practically useful for long-distance communication schemes but would also provide scientific insights into the interaction of light with realistic, turbulent air -- and thus potentially allow for novel measurement techniques of atmospheric effects.

Due to the importance of the question, much effort has been made to study it. To simulate atmospheric turbulence and its effect on complex spatial modes in laboratories, researchers have exploited heat pipes \cite{pors2011transport, chaibi2013propagation}, random phase screens generated by spatial light modulators (SLMs) \cite{rodenburg2012influence, malik2012influence, ibrahim2013orbital,zhang2016experimentally}, static diffractive plates \cite{trichili2016encoding} and rotating random phase plates \cite{ren2013atmospheric}. Those efforts were based on models of the turbulent atmosphere that date back to the Kolmogorov's seminal work from 1941 \cite{kolmogorov1941local, kolmogorov1941degeneration, kolmogorov1941dissipation}, and further extension and advances of it \cite{oboukhov1949structure, corrsin1951spectrum, von1948progress}. 

Kolmogorov's models have since been applied to spatial modes of light (with single phase-screen approximations) \cite{paterson2005atmospheric, tyler2009influence}. Those early methods have been advanced by theories of multi-phase planes developed for complex spatial modes of light \cite{roux2015entanglement} (which have been used in beam propagation simulations already in the late 1980s \cite{martin1988intensity, martin1990simulation}). These models allow for the understanding of spatial light propagation in strong scintillation conditions \cite{mabena2019optical} and comparison of infinitesimal propagation with multi-phase screen propagation \cite{mabena2019corroboration}.

While the theoretical study of atmospheric turbulence effects of spatial modes of light is flourishing, experimental results in real outdoor conditions are lacking. Only in 2014, the first outdoor experiments with spatial modes were conducted, with long-distance transmissions up to 143 km \cite{krenn2014communication, krenn2016twisted}, with high-speed data rates up to 400Gbit/sec \cite{ren2016experimental, li2017high}, and in the quantum regime with entangled photons \cite{krenn2016twisted} and for quantum communication \cite{sit2017high}. These results establish the feasibility of long-distance transmission of spatial modes of light but left open the question about the predictive power of current models. It was only in 2017 when Lavery et al. performed an experiment transmitting spatially modulated light in an urban environment and compared their results with theoretical models \cite{lavery2017free}.

One reason for lack of experimental studies in real-world spatial mode propagation is the uncontrollable environment and the additional noise sources which do not exist in laboratories, such as background stray light, humidity and light absorption. Those are particularly hazardous for quantum experiments. Therefore, as steps to overcome the theoretical-experimental gap, experiments with small complexity are favourable.

Here we explain a detailed numerical study of higher-order spatial modes in different bases propagating through turbulent atmosphere, relying on the method of \cite{lavery2017free, lavery2018vortex}. We find an at first sight effect -- the basis-dependence of atmospheric influence, which could be detected with already existing optical setups. That means information transmitted in one basis (for instance, the famous Laguerre-Gauss (LG) basis carrying orbital angular momentum) is stronger degraded as if the information would have been encoded in another complete basis (such as the Hermite-Gauss (HG) basis). This is interesting to us, as those bases span the same space of modes, and each element of the first basis can be seen as a superposition of elements of the other basis. We find this effect by studying a continuous space of states, the so-called Ince-Gauss (IG) modes \cite{bandres2004ince}, which have both the LG as well as the HG basis as a special case. The predicted effect can be up to 7\% of the total transmission quality; thus, it should be observable in already existing transmission links. If such an effect cannot be observed, it would raise serious questions about some of the best models to describe spatially modulated light in the atmosphere. If the effect is indeed physical, it will shine light on a curious property of the atmosphere and could indicate a novel technique to measure atmospheric properties.

Generally speaking, we have made several contributions in the manuscript: 1). For the first time, we detailed the investigation of Ince-Gaussian modes in turbulence (as a continuously tunable, discrete, complete orthogonal basis which has LG and HG as special cases). 2). As a result of the continuous tunable basis, we identify very concrete basis-dependent effects that have not been reported in the literature before. 3). We present an experimental implementation that will potentially allow for real-world investigations. Our numerically discovered effect could be measured using strong lasers and thus, does not require single-photon or entanglement investigations which are highly elaborate. 4). The number of numerical predictions, due to Ince-Gauss basis's continuous parameter $\epsilon$ is large, and the number of fitting parameters are very small, which results in an ideal target for testing theoretical models accuracy and potential limits. 5). Motivated by the results presented from Lavery et al. \cite{lavery2017free, lavery2018vortex}, we provide potentially interesting experimental observations in the language that requires limited technical vocabulary perfectly suited for experimentalists.

The article is structured in the following way. We start by explaining what we mean by basis-dependent effect and why this is interesting -- especially from a quantum mechanical standpoint in Sec. 2. Then we continue showing how it occurs using light propagation. Details about spatial modes of light are given in Sec. 3. In Sec. 4, we describe the atmospheric turbulence model used for our investigation. In Sec. 5, the numerical results and discussions are shown to illustrate the effect found in our simulation, and we explain a feasible experiment to test this prediction. The conclusions of the paper are given in Sec. 6.

\section{Intuitive description of basis-dependence}
First, we will give a simplified, intuitive explanation of what we call by basis-dependent effects. Our description will be independent of any atmospheric effects and purely motivated by quantum information considerations.

Let us consider a simple two-dimensional sub-basis in a large Hilbert space, which we use to encode information. Let's assume these basis modes undergo a transformation which introduces unbiased loss (for example, scattering into other modes):
\begin{align}
\ket{0}\to t \ket{0} + l \ket{L},\nonumber\\
\ket{1}\to t \ket{1} + l' \ket{L'}
\label{equ:basis1}
\end{align} 
where $t$ stands for a transmission coefficient and $l$ stands for a loss transmission. The fidelity of transmission becomes $F_{\ket{0}}=|t|^2=F_{\ket{1}}$, and its average $F_{0/1}=(F_{\ket{0}}+F_{\ket{1}})/2=|t|^2$, with $\ket{L}$ and its primed version being arbitrary loss modes.

\begin{figure*}[!t]
	\centering
	\includegraphics[width=\linewidth]{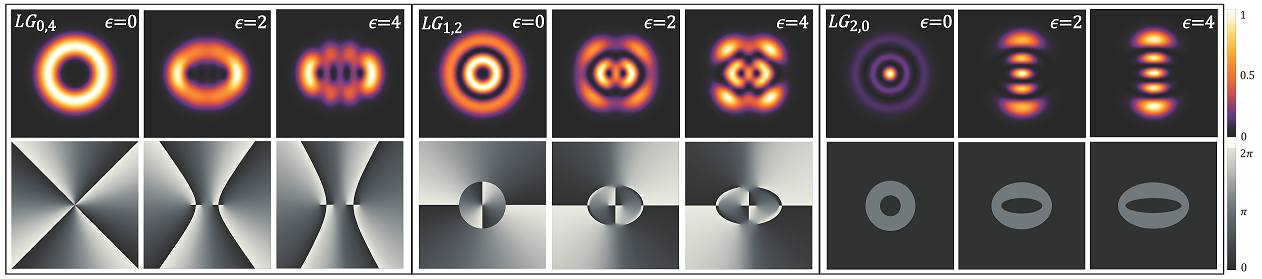}
	\caption{Intensity and phase distributions for IG modes $IG_{p,m,\epsilon}$ with different ellipticity $\epsilon$. A IG mode $IG_{p, m, \epsilon}$ ($\epsilon \rightarrow 0$) is equivalent to a LG mode $LG_{n,\ell}$, where $p=2n+|\ell|$ and $m=|\ell|$. For example, the $LG_{1,2}$ mode can be described as a $IG_{4,2,\epsilon}$ in the limit case of $\epsilon \rightarrow 0$. The upper rows describe the intensity distributions and the lower rows show the transverse phase distributions. A prominent feature of IG modes is the vortex splitting in the phase pattern, leading to multiple intensity nulls in the intensity which are controlled by $\epsilon$.}
	\label{fig:LGIGBeam}
\end{figure*}

However, we could have chosen to encode the information in a different basis, such as the X-basis (with $\ket{\pm}=1/\sqrt{2}\left(\ket{0}\pm\ket{1}\right)$). Using Eq.(\ref{equ:basis1}), the transformation of the $\ket{\pm}$ leads to
\begin{align}
\ket{+}\to t \ket{+} + l'' \ket{L''},\nonumber\\
\ket{-}\to t \ket{-} + l''' \ket{L'''}.
\label{equ:basis2}
\end{align} 
We can immediately see that $F_{\ket{+}}=|t|^2=F_{\ket{-}}$, and its average $F_{D/A}=(F_{\ket{+}}+F_{\ket{-}})/2=|t|^2$, which is exactly the same as the transmission quality in the computation basis $F_{0/1}$. Thus, such transformations are independent of the basis. Experimentally, a \textit{neutral} density filter acts in this way.

Now we give two simple examples of basis-dependence. Consider the transformation, which is a basis-dependent rotation, such as performed by a half-wave plate in polarization optics,
\begin{align}
\ket{0}\to \ket{0},\nonumber\\
\ket{1}\to e^{i \phi} \ket{1}
\label{equ:basis3}
\end{align} 
It leaves the information encoded in the computation basis invariant, $F_{\ket{0}}=1=F_{\ket{1}}$. However, the fidelity in the superposition basis is reduced, $F_{\ket{+}}=|(1+e^{i\phi})/2|^2=F_{\ket{-}}$, thus the transformation leads to basis-dependent transmission qualities.

Let's consider a third example, a mode-dependent loss. 
\begin{align}
\ket{0}\to t\ket{0}+l\ket{L},\nonumber\\
\ket{1}\to (t-\omega)\ket{1}+l'\ket{L'}
\label{equ:basis4}
\end{align} 
The state fidelity after the transmission is $F_{\ket{0}}=|t|^2$, $F_{\ket{1}}=|t-\omega|^2$. We find that in the superposition basis, $F_{\ket{+}}=|t-\omega/2|^2=F_{\ket{-}}$. The difference of the average fidelity of the two bases is $\Delta F=F_{0/1}-F_{D/A}=|\omega|^2/2$, which clearly indicates that the transmission quality of different bases are different, even though they span the same space.

We will see that the atmospheric influence on spatial modes is basis-dependent. As we perform a numerical study, we have to restrict ourselves to a subset of modes. We will see that the influence is mode-dependent when we consider a complete orthogonal \textit{subspace} of modes with the same mode-order, in Table \ref{tab:fidelityOrder}. This is particularly interesting, as the mode-order is defined independent of the basis, and a specific mode of one basis (with order $M$) can be decomposed in a superposition of modes of order $M$ in arbitrary other bases. Of course, the considerations here are vast simplifications of influence of the atmosphere on spatial modes, and can only be considered as a simple, intuitive picture of how to understand such effects in general.

\section{Complex spatial modes of light}

The well-known solutions to the paraxial wave equation consist of the HG and LG beams, which are derived from cartesian and circular cylindrical coordinates, respectively \cite{siegman1986lasers}. They both provide in principle an infinite state space and form a complete orthogonal basis, such that one can describe HG states in terms of LG modes and vice versa \cite{kimel1993relations, beijersbergen1993astigmatic}. The HG modes are denoted as $HG_{n_{x},n_{y}}$ with indices $n_{x}$ and $n_{y}$ and LG modes are described as $LG_{n,\ell}$ with orbital angular momentum (OAM) index $\ell$ \cite{allen1992orbital} and radial number $n$ \cite{karimi2012radial, karimi2014radial, plick2015physical}. 

In 2004, Bandres and Guti\'{e}rrez-Vega introduced another complete and orthogonal family of modes, which interpolates between the LG and HG modes  \cite{bandres2004ince}. These modes are exact solutions to the paraxial wave equation in elliptic coordinates, and of which the even IG modes are described as
\begin{align}
IG&_{p,m,\epsilon}^{e}(u,v,z)=\frac{C_{IG}w_{0}}{w_{z}}C_{p}^{m}(iu,\epsilon)C_{p}^{m}(v,\epsilon)\\\nonumber
&\times \exp\left(-\frac{r^{2}}{w_{z}^{2}}+i\left(\frac{kr^{2}}{2R_{z}}-kz-(p+1)\varphi_{g} \right) \right).
\label{equ:IGbeam}
\end{align}
There $u$ and $v$ describe the two-dimensional elliptic coordinates. A continuous parameter $\epsilon$ describes the ellipticity and the superscript $e$ refers to even modes. In the limit case of $\epsilon \rightarrow 0$, $u$ and $v$ correspond to the radial and angular coordinates of the circular cylindrical coordinates system, respectively. $(p,m)\in\mathbb{N}$ are mode number. $C_{IG}$ is a normalization constant and $C_{p}^{m}(\cdot,\epsilon)$ represent the even Ince polynomials \cite{arscott2014periodic, bandres2014}. $w_{z}$ is the beam radius at position $z$ and $w_0$ is the beam waist at the focus $z=0$. $z_{R}$ is the Rayleigh range, $R_{z}$ is the radius of curvature, $\lambda$ is the wavelength and $k=2\pi/\lambda$ is the wave number. $\varphi_{g}=\arctan(z/z_{R})$ denotes the Gouy phase and the mode order is $M=p$. We can obtain the odd IG modes $IG_{p,m,\epsilon}^{o}(u,v,z)$ by replacing the $C_{IG}$ and $C_{p}^{m}(\cdot,\epsilon)$ with $S_{IG}$ and $S_{p}^{m}(\cdot,\epsilon)$, which correspond to a normalization constant and the odd Ince polynomials respectively.

\begin{figure*}[!t]
	\centering
	\includegraphics[width=0.95\linewidth]{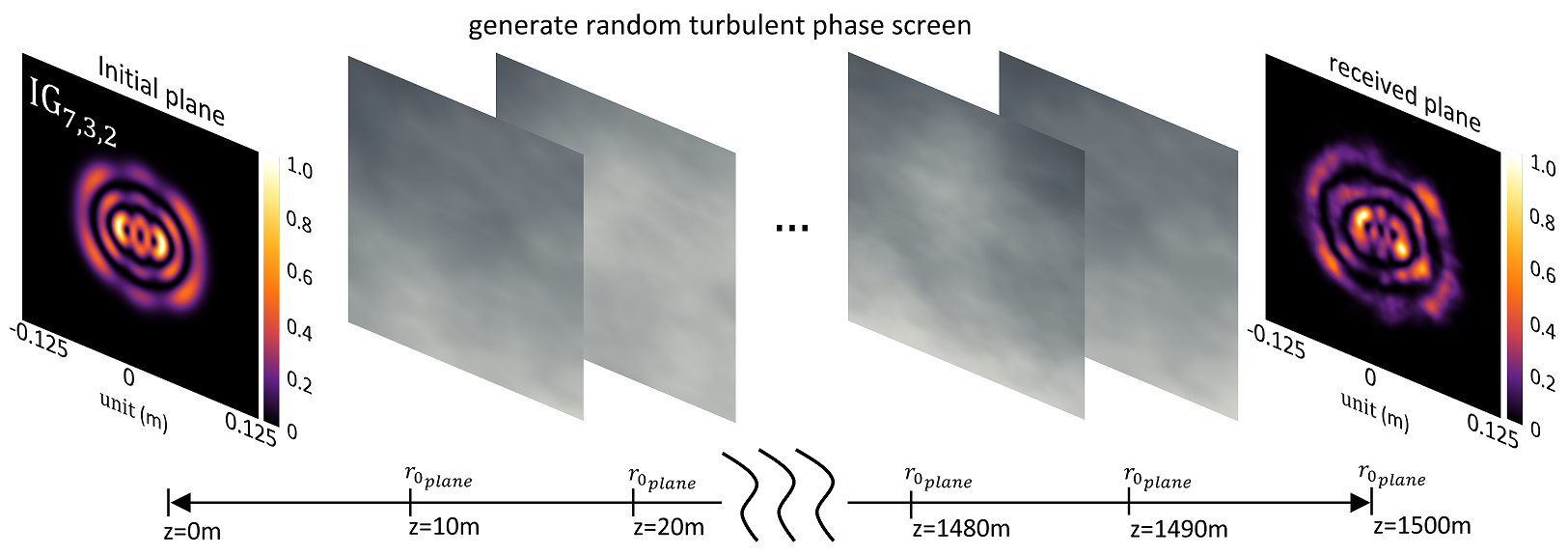}
	\caption{A schematic illustration of spatial modes of light propagating through atmospheric turbulence. Here we transmit a IG mode $IG_{7, 3, 2}$ ($\lambda=809$nm, $w_{0}=25$mm ) over a 1.5 km turbulent free-space link. The intensity distributions of the IG mode $IG_{7, 3, 2}$ in the transmitted and received planes are shown. We use the multi phase-screen method, detailed theoretically in \cite{roux2015entanglement}, and used to describe empirically observed effects in \cite{lavery2017free}, to model the overall turbulent link. There the total effect of atmospheric turbulence can be represented as the approximately accumulated influence of many weakly perturbing planes. Each plane with a random phase screen stands for the turbulence along a propagation path of 10 m, and the scintillation strength of each plane is described by the Fried parameter $r_{0_{plane}}$. In general, larger $r_{0_{plane}}$ defines weaker turbulence and in this example the number is $r_{0_{plane}}=1$ m (corresponding to $r_{0}\approx 0.05$ m).}
	\label{fig:simulation}
\end{figure*}

The helical IG modes (which, for simplicity, we call IG modes for the rest of the paper) are described as superposition of even and odd IG modes \cite{bandres2004inceosa, bentley2006generation}, which are given by
\begin{align}
IG&_{p,m,\epsilon}^{\pm}(u,v,z)\\\nonumber
&=\frac{1}{\sqrt{2}}\left(IG_{p,m,\epsilon}^{e}(u,v,z)\pm iIG_{p,m,\epsilon}^{o}(u,v,z)\right).
\end{align}

When the elliptic coordinates tend to the circular cylindrical coordinates, namely $\epsilon \rightarrow 0$, the IG modes will be transferred into LG modes. In this case, the indices of states $LG_{n,\ell}$ and $IG_{p,m,\epsilon}$ are related as: $|\ell|=m$ and $n=(p-m)/2$. Additionally, when the elliptic coordinates tend to the cartesian coordinates, namely $\epsilon\rightarrow\infty$, the IG modes will be transferred into helical Hermite-Gaussian (HG) modes \cite{bandres2004inceosa, lopez2007propagation, plick2013quantum}. In this case, the indices of modes $HG_{n_{x},n_{y}}$ and $IG_{p,m,\epsilon}$ are related as: $n_{x}=m$ and $n_{y}=(p-m)$. This puts the IG modes in a special position between LG and HG, and therefore makes them an ideal workhorse for investigating basis-dependent effects.

For a fixed ellipticity $\epsilon$, the IG modes $IG_{p,m,\epsilon}$ with orthogonal mode indices $(p,m)\in\mathbb{N}$ form a complete basis. Changing the value of the ellipticity $\epsilon$ gives another complete orthogonal basis. In Fig. \ref{fig:LGIGBeam}, we show the intensity and phase distributions of different IG modes $IG_{p,m,\epsilon}$ varying their ellipticity $\epsilon$.

There are also other well-investigated types of complete sets of transverse spatial modes. One example is the family of Bessel modes which posses intrinsic self-healing properties that could be of interest in out-door turbulence experiments \cite{mclaren2014self}, or their elliptic generalizations called Mathieu beams \cite{gutierrez2000alternative,gutierrez2001experimental}. However, here we focus on the Ince-Gaussian basis.

\section{Turbulence model}

The inhomogeneity and anisotropy in the temperature and pressure of atmosphere results in random fluctuations of the refractive index along the propagation path of light beam \cite{andrews2005laser}. Those variations of refractive index introduce the distortions of the spatially structured light fields and increase intermodal crosstalk, which dramatically affects the quality of spatial modes over long-distance link \cite{li2018atmospheric}. Thus understanding the turbulent behaviour of the atmosphere is very crucial.

In 1941, the Russian mathematician Andrey Kolmogorov published three seminal articles, which established the foundations of statistical turbulence theory \cite{kolmogorov1941local, kolmogorov1941degeneration, kolmogorov1941dissipation}. The statistical properties of the atmospheric turbulence are homogeneous and isotropic within scale $L_0$ and $l_0$.  He found that the random statistical behaviour of turbulence can be described by refractive index power density spectrum, which describes how the kinetic energy of atmospheric turbulence distributes with respect to frequencies. This directly relates to the phase fluctuations of light along propagation path and later is represented by the phase power density spectrum $\Phi_{\varphi}^{mvK}(\kappa)$ \cite{andrews2005laser}. In our study, we apply the modified von K\'{a}rm\'{a}n spectrum \cite{von1948progress}, which avoids the singularity that represents energy per unit volume and becomes unbounded as the eddy size increases in Kolmogorov spectrum model. Therefore it is numerically more stable. The turbulent phase screens are generated using the modified von K\'{a}rm\'{a}n spectrum, which is described as \cite{schmidt2010numerical, osborn2010profiling}
\begin{align}
\Phi_{\varphi}^{mvK}(\kappa)=0.023 r_{0}^{-5/3}\frac{exp(-\kappa^{2}/\kappa_{m}^{2})}{({\kappa^{2}+\kappa_{0}^{2}})^{11/6}}.
\end{align}

There $r_{0}$ is the Fried parameter \cite{fried1965statistics}, which can be used to describe the strength of the atmospheric turbulence along the propagation distance. $\kappa$ is the angular spatial frequency. $\kappa_{m}$ and $\kappa_{0}$ are constant model-dependent parameters \cite{andrews2005laser, schmidt2010numerical, osborn2010profiling}.

In order to simulate the atmospheric turbulence along a long propagation distance, we adopt a multi-phase plane model \cite{roux2015entanglement,lavery2018vortex}. In the model shown in Fig. \ref{fig:simulation}, the atmosphere along 1.5 km link is split into 150 planes of weakly turbulence separated by 10 m. The total atmospheric turbulence $r_{0}$ is approximately the accumulative effect of the turbulence of each plane $r_{0_{plane}}$. In our numerical simulation, we satisfy geometric and aliasing constraints \cite{schmidt2010numerical, coy2005choosing, tang2015propagation}.

\subsection{Turbulence strength and Model}
In the presence of a turbulent atmosphere, spatial modes of light experience atmospheric refractive index variations caused by fluctuations in temperature and pressure. These atmospheric refractive index variations distort the wavefronts of the propagated light beams \cite{andrews2005laser}. A measure of the strength of random fluctuations is the refractive index structure parameter $C_{n}^2(z)$, which is used to quantify the strength of the atmospheric turbulence along the propagation path. Typical values of $C_{n}^2(z)$ range from $10^{-17}$ $m^{-2/3}$ in weak scintillation up to $10^{-13}$ $m^{-2/3}$ in strong turbulence \cite{andrews2005laser, osborn2010profiling}.

Another parameter often used to estimate the integrated strength of turbulence,  especially in connection with astronomical imaging, is the Fried parameter $r_{0}$ \cite{fried1966optical}. Stronger turbulence corresponds to a smaller $r_{0}$. For a known refractive index structure parameter $C_{n}^2(z)$ along the propagation path, the Fried parameter $r_{0}$ is given by \cite{roddier1981v, sasiela2007electromagnetic, schmitt2007long}
\begin{equation}
r_{0}=\left(\alpha_{1} k^2\int_{path}C_{n}^2(z)dz\right)^{-3/5}.
\label{eq:friednumber}
\end{equation}

There $k=2\pi/\lambda$ is the wavenumber and $\lambda$ is optical beam wavelength (in our simulations, a wavelength of $809nm$ is adopted). $\alpha_{1}=0.423$ is a constant number which derived in the case of the phase variance is approximately one \cite{roddier1981v, osborn2010profiling}. The integral is taken over the overall propagation path from the transmitter plane to the receiver plane.

We rely on the experimentally demonstrated model from Lavery et al. \cite{lavery2017free, lavery2018vortex} to simulate the atmosphere over long-distance. There the total 1.5 km turbulent link is decomposed into many short weakly perturbing planes separated by 10 m. In general, $C_{n}^2(z)$ is assumed to be roughly a constant over short time intervals or propagation distance. Using Eq. \ref{eq:friednumber}, we could approximately describe the overall strength of atmosphere $r_{0}$ as an accumulation of strength in every plane $r_{0_{plane}}$, which is described as
\begin{equation}
r_{0}^{-5/3}\approx\frac{z}{\Delta z} r_{0_{plane}}^{-5/3}=150r_{0_{plane}}^{-5/3}.
\label{eq:friedplane}
\end{equation}

The values of different turbulence conditions used for studying the propagation of spatial modes of light are listed in Table. \ref{tab:TurbulenceStrength}.

For the purpose of numerically modeling turbulence, we use the modified von K\'{a}rm\'{a}n phase power spectrum $\Phi_{\varphi}^{mvK}(\kappa)$ to generate random phase screens \cite{von1948progress, osborn2010profiling}, which is described as
\begin{equation}
\Phi_{\varphi}^{mvK}(\kappa)=\alpha_{2}k\int_{path}C_{n}^2(z)dz \frac{exp(-\kappa^{2}/\kappa_{m}^{2})}{({\kappa^{2}+\kappa_{0}^{2}})^{11/6}}.
\label{eq:models}
\end{equation}
where $\kappa$ is angular spatial frequency in rad/m. $\alpha_{2}=9.7\times10^{-3}$, $\kappa_{m}=5.92/l_{0}$ and $\kappa_{0}=2\pi/L_{0}$ are constant model-dependent parameters \cite{osborn2010profiling, schmidt2010numerical}. $L_{0}$ and $l_{0}$ are the so-called outer and inner scale, which describe the averaged largest and smallest eddies for the kinetic energy distribution in atmospheric turbulence. Their typical value are $L_{0}=100$ m and $l_{0}=0.01$ m \cite{schmidt2010numerical}. The phase power spectrum $\Phi_{\varphi}^{mvK}(\kappa)$ can also be written in terms of a Fried parameter $r_{0}$ by combing Eqs. \ref{eq:friednumber} and \ref{eq:models}, which is given as
\begin{equation}
\Phi_{\varphi}^{mvK}(\kappa)=0.023 r_{0}^{-5/3}\frac{exp(-\kappa^{2}/\kappa_{m}^{2})}{({\kappa^{2}+\kappa_{0}^{2}})^{11/6}}.
\label{eq:Friedmodels}
\end{equation}

\begin{table}[!t]
	\centering
	\caption{Strengths of atmospheric scintillation.}        
	\begin{tabular}{|c|c|c|c|c|c|c|}
		\hline
		$r_{0_{plane}}$ /m& 0.25 & 0.55 & 1.0 & 1.5 & 2.0 & 2.5 \\ \hline
		$r_{0}$ /m & 0.012 & 0.027& 0.049 & 0.074 & 0.099 & 0.124 \\ \hline
		\multirow{2}{*}{$C_{n}^2$ /$m^{-2/3}$} & \multicolumn{2}{c|}{$>10^{-14}$} & \multicolumn{4}{c|}{$10^{-16}\sim10^{-14}$} \\ 
		& \multicolumn{2}{c|}{Strong} & \multicolumn{4}{c|}{Moderate}\\ \hline
	\end{tabular}
	\label{tab:TurbulenceStrength}
\end{table}

\subsection{Numerical Simulation}
In our simulation, we adopted a collimated light beam of beam waist $w_{0}=25mm$ and wavelength $\lambda=809nm$. All parameters used in our numerical simulation are presented in Table. \ref{tab:SimultatedParameters}.
\begin{table}[ht]
	\centering
	\caption{Parameters used in our numerical simulation.}
	\begin{tabular}{|c|c|c|}
		\hline
		Parameters & Values & Units \\ \hline
		optical wavelength $\lambda$ & 809 & $nm$ \\ \hline
		optical beam waist $w_{0}$ & 0.025& $m$ \\ \hline
		propagation distance $z$& 1500& $m$ \\ \hline
		distance interval $\Delta z$& 10& $m$ \\ \hline      
		scintillation strength $r_{0}$ & 0.05 & $m$ \\ \hline
		outer scale $L_{0}$ & 100 & $m$  \\ \hline
		inner scale  $l_{0}$& 0.01 & $m$  \\ \hline
		number of turbulent planes $n$& 150 & -\\ \hline
		number of grids $N$& 1024& -\\ \hline
		side length at source plane $L_{1}$ & 0.25& $m$ \\ \hline
		side length at receiver plane $L_{n}$ & 0.5& $m$ \\ \hline
	\end{tabular}
	\label{tab:SimultatedParameters}
\end{table}

Then we investigate the propagation quality of IG modes $IG_{p,m,\epsilon}$ with different ellipticity under different turbulence conditions. The transmission fidelity describes the "closeness" of the received propagated light field $\ket{\Psi_{tur}}$ and the undisturbed light field in the observed planes $\ket{\Psi_{vac}}$, which is given by the squared overlapping the two light fields as
\begin{equation}
F=|\braket{\Psi_{vac}|\Psi_{tur}}|^{2}
\label{eq:fidelity}
\end{equation}

The average fidelity is computed by averaging all individual transmissions. For the average fidelity of certain mode order $M$, it is given by over all the modes in the same order and basis. Furthermore, we investigate our finding by analyzing the mode cross-talk matrix for complex spatial modes $IG_{p,m,\epsilon}$ with different ellipticity $\epsilon$. In the limit of $\epsilon\rightarrow0$, a chosen basis with IG indices set $(p,m)$ can be rewritten as a LG indices $(n,\ell)$. Thus, in the cross-talk matrix labeled columns represent the modes with index $\ell$, meanwhile labeled rows represent the modes with index $n$. Each element in the matrix is given by an inner product measurement of the transmitted field $\ket{\Psi_{tur}}$ and each undisturbed mode (individually) from the set of basis $\ket{vac_{j}}$, which is given by
\begin{equation}
F_{j}=|\braket{\Psi_{vac_{j}}|\Psi_{tur}}|^{2}.
\label{eq:fidelitymatrix}
\end{equation}

\section{Results and discussion}
What is the quality of spatial modes of light propagating through atmospheric turbulence over long-distance? We start with transmitting an IG mode $IG_{4,0,\epsilon}$ over a 1.5 km turbulent channel. We call radial-like modes for IG states $IG_{p,m,\epsilon}$ when they are equivalent to radial modes $LG_{n,0}$ in the limit case of $\epsilon\rightarrow0$. Analogously, we call OAM-like modes for IG states $IG_{p,m,\epsilon}$ when they are equivalent to OAM modes $LG_{0,\ell}$ in the limit case of $\epsilon\rightarrow0$. Therefore, the IG mode $IG_{4,0,\epsilon}$ used in our simulation corresponds to a radial-like mode, with two intensity rings with ellipticity $\epsilon=0$.

\begin{figure}[!t]
	\centering
	\includegraphics[width=1\linewidth]{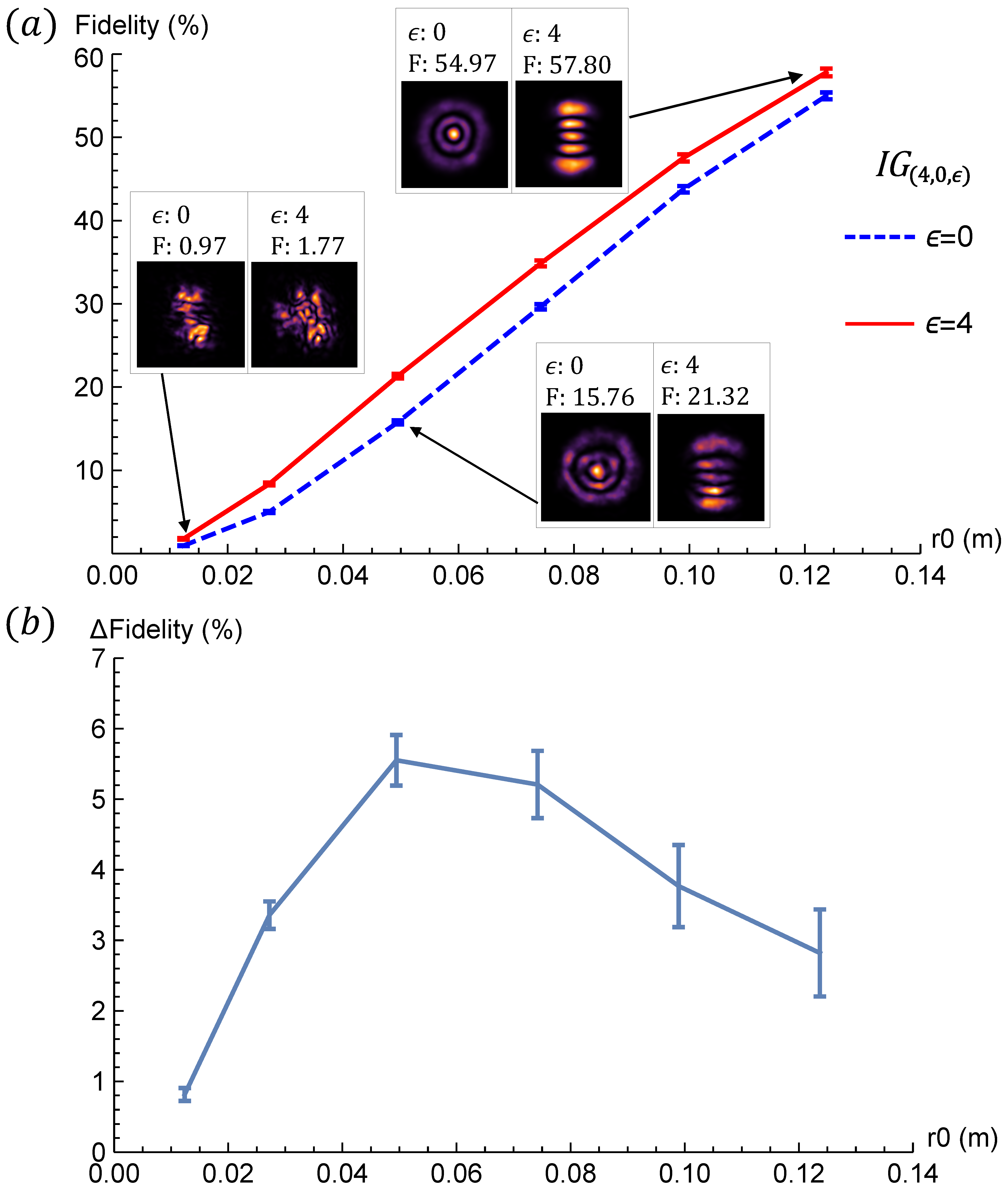}
	\caption{The propagation quality of IG modes under different turbulence conditions. \textbf{a}: We transmit a radial-like mode $IG_{4,0,\epsilon}$ ($\epsilon:0,4$) over 1.5 km turbulent link of different strengths $r_{0}$. Large $r_{0}$ describes weak scintillation and indicates good transmission fidelity. We find that the quality of IG modes in the case of $\epsilon=4$ is significantly better than that in the case of $\epsilon \rightarrow0$. \textbf{b}: We show the fidelity difference versus turbulent strength $r_{0}$. There is a large difference when the turbulent strength is $r_{0}\approx 0.05$ m. If we continue to increase $r_{0}$ (which means turbulence becomes weak until there is no turbulence), the difference will vanish.}
	\label{fig:Strengthresults}
\end{figure}

Here a question naturally arises: What role does the ellipticity play on the transmission quality of IG modes under different turbulent conditions? For simplicity, we use two different ellipticity $\epsilon\rightarrow0$ and $\epsilon=4$ and propagate the IG modes through the atmosphere of different turbulent strength $r_{0}$. The average fidelity of $IG_{4,0,\epsilon}$ through different turbulent strengths are described in Table. \ref{tab:Fidelityturbulence} and Fig. \ref{fig:Strengthresults} (a). There the fidelity is in percentage and the error is given by the standard deviation of the mean.
\begin{table}[!t]
	\centering
	\caption{The propagation fidelity of $IG_{4,0,\epsilon}$ through different turbulent conditions $r_{0}$. The iterations are not round numbers as they have been stopped (because of large computational costs) when all results where statistically significant.}
	\begin{tabular}{|c|c|c|c|c|}
		\hline
		\multirow{2}{*}{iteration} & \multirow{2}{*}{$r_{0} (m)$} & \multicolumn{2}{c|}{average fidelity $F$ (in $\%$)} & \multicolumn{1}{c|}{$\triangle F_{IG-LG}$} \\
		\cline{3-4}
		& & IG: $\epsilon\rightarrow0$ & IG: $\epsilon=4$ & (significance: $\sigma$)\\ \hline
		670 & 0.012& 0.95$\pm$0.04  & 1.77$\pm$0.08 & 0.81$\pm$0.09 (9$\sigma$) \\ \hline
		3136& 0.027& 5.00$\pm$0.11  & 8.36$\pm$0.16 & 3.36$\pm$0.19 (17$\sigma$)\\ \hline
		4640& 0.049& 15.77$\pm$0.23  & 21.32$\pm$0.27 & 5.55$\pm$0.36 (15$\sigma$)\\ \hline
		4828& 0.074& 29.65$\pm$0.31  & 34.86$\pm$0.36 & 5.21$\pm$0.48 (11$\sigma$) \\ \hline
		3798& 0.099& 43.76$\pm$0.38  & 47.53$\pm$0.44 & 3.77$\pm$0.58 (6.5$\sigma$) \\ \hline
		3112& 0.124& 54.98$\pm$0.41  & 57.80$\pm$0.47 & 2.82$\pm$0.62 (4.5$\sigma$) \\ \hline
	\end{tabular}
	\label{tab:Fidelityturbulence}
\end{table}

The result clearly shows the fact that under weak scintillation (equivalent to the case of large $r_{0}$), the quality is better than that under strong turbulence. Interestingly, we find that the ellipticity plays an interesting role in the transmission quality of IG modes along the turbulent path. To our surprise, there is a significant increase for the radial-like modes with a large ellipticity propagating through strong turbulence. In Fig. \ref{fig:Strengthresults} (b), we show the fidelity difference versus various turbulent strengths. We find that there is a large fidelity difference when the turbulent strength is $r_{0}\approx0.05$ m, which is a realistic turbulence condition (we use this $r_{0}$ for the rest of our simulations). With this observation, we would expect that the difference will continue to increase by enlarging the ellipticity of the radial-like mode $IG_{4,0,\epsilon}$. This indicates that helical HG modes, corresponding to radial-like modes in the case of $\epsilon \rightarrow \infty$, perform better under strong turbulence than that by LG radial modes.
\begin{table}[!t]
	\centering
	\caption{The transmission fidelity of $IG_{4,0,\epsilon}$ with different $\epsilon$ through turbulence with strength $r0\approx 0.05$ m.}
	\begin{tabular}{|c|c|c|c|}
		\hline
		\multirow{2}{*}{$IG_{p,m,\epsilon}$} & \multicolumn{2}{c|}{average fidelity $F$ (in $\%$)} & \multicolumn{1}{c|}{$\triangle F_{IG-LG}$} \\
		\cline{2-3}
		& IG: $\epsilon\rightarrow0$ & IG: $\epsilon=4$ & (significance: $\sigma$)\\ \hline
		$IG_{4,0,\epsilon}$& 14.23$\pm$0.46  & 20.37$\pm$0.56 & 6.15$\pm$0.73 (8.4$\sigma$) \\ \hline
		$IG_{4,2,\epsilon}$& 14.18$\pm$0.48  & 14.64$\pm$0.49 & 0.45$\pm$0.68 (0.5$\sigma$)\\ \hline
		$IG_{4,4,\epsilon}$& 15.87$\pm$0.52  & 16.88$\pm$0.53 & 1.01$\pm$0.74 (1.3$\sigma$)\\ \hline
	\end{tabular}
	\label{tab:TwoIGradialFidelity}
\end{table}

\begin{figure}[!t]
	\centering
	\includegraphics[width=\linewidth]{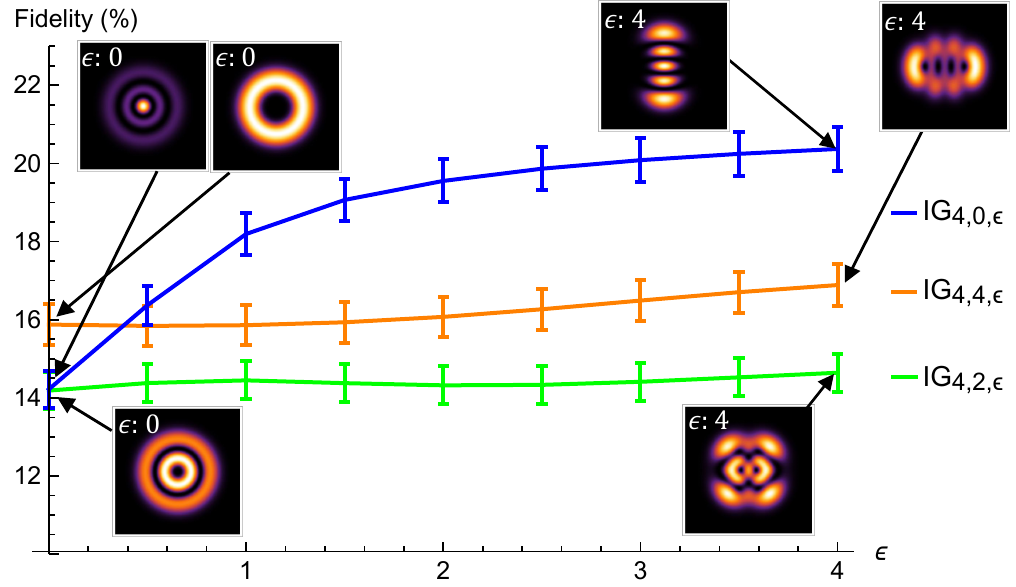}
	\caption{The propagation fidelity of IG modes with different ellipticity through atmospheric turbulence. Here we analyse a set of states of mode order $M=4$ \{$IG_{4,0,\epsilon}$, $IG_{4,4,\epsilon}$, $IG_{4,2,\epsilon}$\} with 9 different $\epsilon$. In the case of $\epsilon \rightarrow$ 0, such a IG mode set is equivalent to the LG mode set \{$LG_{2,0}$, $LG_{0,4}$, $LG_{1,2}$\}. The insets describe the theoretical intensities of these IG modes ($\epsilon:0,4$). The result shows that, with a large $\epsilon$, there is a significant increase in the fidelity of radial-like modes and there is no decrease in the fidelity of OAM-like modes.}
	\label{fig:Singlemoderesults}
\end{figure}
The transmission quality for radial-like modes increases when we increase the ellipticity $\epsilon$. We would therefor expect that this increase is compensated by OAM-like modes whose quality decreases when we increase $\epsilon$, such that the transmission quality for a complete order of modes stay constant. However, the results in Table. \ref{tab:TwoIGradialFidelity} and Fig.\ref{fig:Singlemoderesults} show that the transmission quality of all other modes with order $M=4$ remains constant (within significant uncertainty). This means that the average propagation quality of order $M=4$ increases when we change to a basis with a large ellipticity. This means that the atmosphere influences bases with small $\epsilon$ in a stronger way than bases with large $\epsilon$. Every mode with order $M$ of a specific basis can be decomposed into a coherent sum of modes of order $M$ in another basis. For that reason, observing the effect in order $M=4$ is sufficient to conclude a basis-dependent effect. This is the basis-dependent effect of atmospheric turbulence we present here, and its investigation is the main results of our paper.

\begin{figure}[!t]
	\centering
	\includegraphics[width=\linewidth]{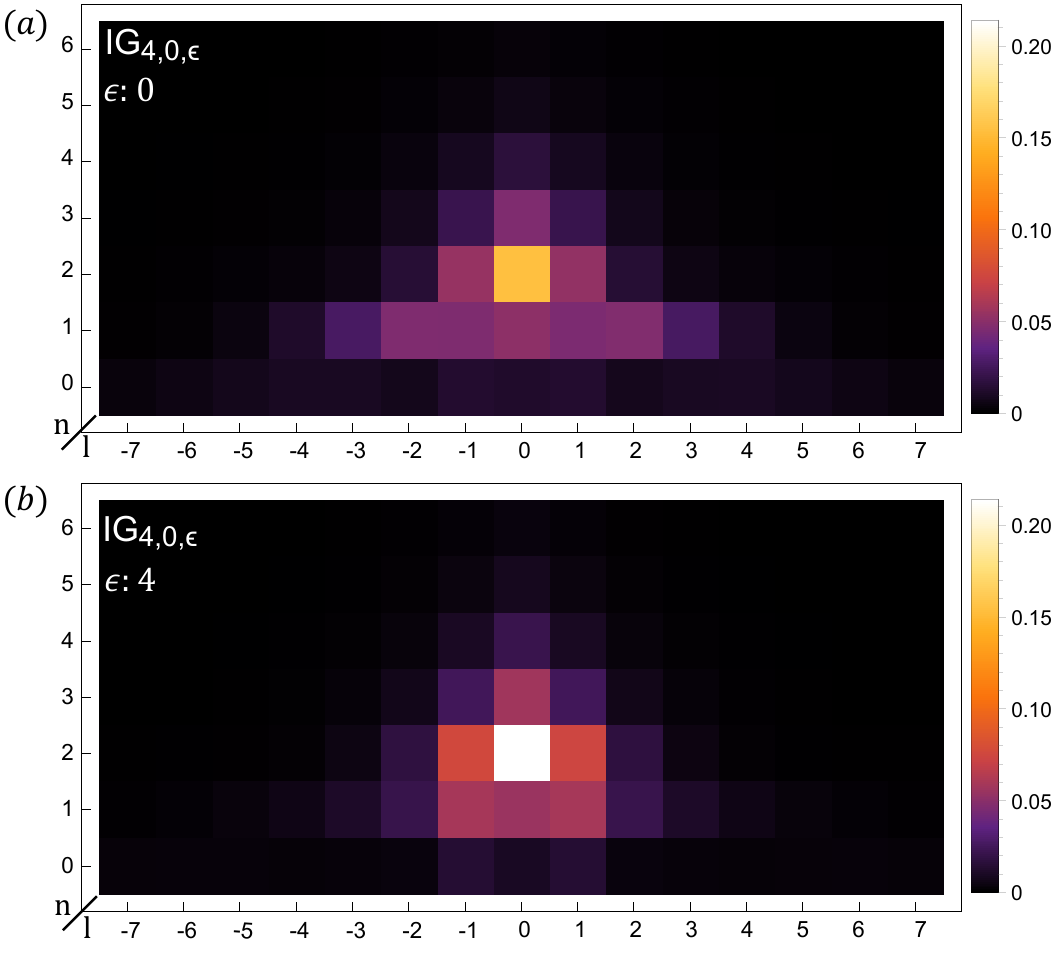}
	\caption{The mode cross-talk matrix for radial-like modes $IG_{4,0,\epsilon}$ propagating through turbulence. In the limit of $\epsilon\rightarrow0$, a chosen basis with IG indices set $(p,m)$ can be rewritten as a LG indices $(n,\ell)$, where $p=2n+|\ell|$ and $m=|\ell|$ and in our example $\ell\in$ (-7,7) and $n\in$ (0,6). Each element in the matrix stands for the fidelity between the mode after transmission of 1.5 km, and an undisturbed mode. In a vacuum, only one element would have value F=1, and everything else would be zero. \textbf{a}: The mode cross-talk matrix for $IG_{p,m,\epsilon}$ with $\epsilon\rightarrow 0$. \textbf{b}: The mode cross-talk matrix for $IG_{p,m,\epsilon}$ with $\epsilon=4$. We can see that in the case $\epsilon\rightarrow0$ (LG modes), the fidelity is spreading over significantly more modes than for $\epsilon=4$. The physical reason for this phenomenon should be a target for a follow-up investigation.}
	\label{fig:Possiblereason}
\end{figure}

Furthermore, we find that this effect consistently exists for order $M=2$ up to $M=6$ (which contains 3 and 7 modes, respectively). For mode order $M=0$ and $M=1$, the ellipticity does not change the modes. Therefore, mode order $M=2$ is the smallest that we investigate here. We show the results in Table. \ref{tab:fidelityOrder}. An interesting insight into this effect is a cross-talk matrix of the radial-like mode $IG_{4,0,\epsilon}$ with small and large ellipticity. Indeed, in Fig. \ref{fig:Possiblereason} we observe that small ellipticity leads to larger cross-talk with other modes of this basis, whereas larger ellipticity reduces the cross-talk. In addition, when we rotate the HG modes and $IG_{4,0,\epsilon}$ modes with higher $\epsilon$, such mode-dependence effect still exit, see Appendix A for more details. We have already explained an intuitive way of understanding this mode-dependence in Sec. 2, and a fully mathematical description of this effect would be exciting, but that is out of the scope of this manuscript.
\begin{table*}[!ht]
	\centering
	\caption{Average fidelity for modes with different orders $M$ from a different basis. We calculate all modes in the same order $M$ (each mode with 2000 iterations). The average fidelity is over all modes in the same order $M$ in the same basis, such as order $M=2$ contains 3 different modes in one basis. The fidelity is in percentage, and the standard deviation of the mean gives the error. The result indicates that the atmosphere introduces a basis-dependent effect. An intuitive description is given in Sec. 2.}
	\begin{tabular}{|c|c|c|c|c|c|}
		\hline
		\multirow{2}{*}{order $M$}& \multicolumn{3}{c|}{average fidelity $F$ (in $\%$)} & \multirow{2}{*}{$\triangle F_{IG-LG}$ (significance: $\sigma$)} & \multirow{2}{*}{$\triangle F_{HG-LG}$ (significance: $\sigma$)}\\
		\cline{2-4}
		& LG: IG $\epsilon\rightarrow0$ & IG: $\epsilon=4$ & HG & & \\ \hline
		2& 23.35$\pm$0.27 & 25.23$\pm$0.28  & 26.57$\pm$0.28 & 1.88$\pm$0.38 (5.0$\sigma$)  & 3.22$\pm$0.39 (8$\sigma$) \\ \hline
		3& 18.62$\pm$0.20 & 20.23$\pm$0.21  & 22.05$\pm$0.22 & 1.61$\pm$0.29 (5.5$\sigma$)& 3.43$\pm$0.30 (11$\sigma$) \\ \hline
		4& 15.57$\pm$0.16 & 17.37$\pm$0.17  & 18.99$\pm$0.18 & 1.80$\pm$0.24 (7.5$\sigma$)& 3.42$\pm$0.24 (14$\sigma$) \\ \hline
		5& 13.46$\pm$0.13 & 14.84$\pm$0.14  & 16.75$\pm$0.15 & 1.38$\pm$0.19 (7.1$\sigma$)  & 3.28$\pm$0.20 (16$\sigma$) \\ \hline
		6& 11.90$\pm$0.11 & 13.41$\pm$0.12  & 15.00$\pm$0.13 & 1.51$\pm$0.16 (9.5$\sigma$)& 3.10$\pm$0.17 (18$\sigma$) \\ \hline
	\end{tabular}
	\label{tab:fidelityOrder}
\end{table*}

There is now one important question that remains: Is this a physical effect or an artefact of a currently well-trusted model for the propagation of complex spatial modes? This question can only be solved by experiments. An experimental investigation would require a long-distance outdoor link for transmitting spatial modes of light, which already exists in several places worldwide. At the sender, one requires a high-quality construction of complex spatial modes, for instance, using the technique presented in \cite{bolduc2013exact}. At the receiver, the mode needs to be measured -- for example by transforming it to a Gaussian mode with high-quality \cite{bouchard2018measuring} and using a Single-Mode Fiber as a filter \cite{mair2001entanglement,krenn2013entangled}. Atmospheric conditions (in particular the Fried parameter $r_0$) are stable for long enough to perform measurements in the form of Fig. \ref{fig:Singlemoderesults} and \ref{fig:Possiblereason} successfully. The experiment is well within reach of current technology. It could be one interesting target for detailed investigations of turbulence models using a complex, continuously class of complete, orthogonal spatial mode families. In particular, with merely one free fit-parameters $r_0$ (which could be measured by other means), the experiments can test many predictions, such as fidelities of different modes, such as shown in Fig. \ref{fig:Singlemoderesults} and \ref{fig:Possiblereason}, and in particular, the important quantity of mode-dependence $\Delta F$ in Table. \ref{tab:fidelityOrder} . Other tests could involve varying of beam radii (which we didn't explicitly calculate), but which can be performed in the same setup with only adjusting the computer-generated holograms.

\section{Related work}
Several recent studies have compared properties between Hermite-Gauss modes and OAM modes (a subset of Laguerre-Gaussian modes) in terms of their classical communication quality. In 2016, Restuccia et al. have shown that in real-world communication, the finite-size apertures of sending and detection systems degrade the quality of the communication channel. The authors have shown that in such cases, Laguerre-Gaussian beams are more suitable than Hermite-Gauss modes \cite{restuccia2016comparing}. We do not take finite-size effects of the measurement systems into account as we are interested in the fundamental physical properties of light propagation (and we are -- for the moment -- agnostic to technical implementations).

An alternative result has been obtained by the authors of \cite{ndagano2017comparing}, which has been recently experimentally simulated in the lab \cite{cox2019resilience}. The authors show that a subset of Hermite-Gauss modes is less perturbed than doughnut modes (a subset of Laguerre-Gaussian modes with the radial mode index $n$=0) by single-phase screens for approximating turbulence effects. The effect stems from larger robustness against lateral translation because Hermite-Gauss modes are \textit{azimuthally asymmetric}. Their work, which is essential for applications in classical communications (for instance, see \cite{milione2018using, milione2019free}) differs in several ways from our findings: First, we investigate the full two-dimensional spatial basis of light, which is defined by two-mode numbers of Laguerre-Gaussian mode. This involves the azimuthal part (which the authors examine) and the radial part (which is not investigated, and which is known to have different physical properties, for instance, see \cite{karimi2014radial}). Only the combination of all spatial mode numbers lets us make conclusions about the physical effect of the basis-dependent phenomena. Second, the results obtained by the authors are qualitatively similar (but quantitative different) compared to our findings, and it is not immediately clear why some asymmetric Ince-Gaussian modes (as shown by the orange and green line in Fig. \ref{fig:Singlemoderesults}) do not improve the fidelity when the asymmetry (ellipticity) is increased. Furthermore, we use an empirically suggested model of light propagation through turbulence because single-phase approximations cannot cover all experimentally observed effects \cite{lavery2017free}. We would argue that the best way to understand this effect is actually to perform the real-world experiment.

A different, fascinating recent work is conceptually related to our investigation: The authors investigate the propagation of doughnut modes (Laguerre-Gaussian beam with radial mode number $n$=0) through simulated turbulence. They present that for an OAM mode under realistic turbulence conditions, the spiral spectrum develops two maxima. One of the maxima is on the negative axis, and one is on the positive axis of the spectrum, while OAM=0 is very unlikely to appear \cite{sorelli2019entanglement}. It would be exciting to observe the predicted effect in a real-world experiment.

The mode-dependent influence of turbulence has been seen in theoretical calculations several times, for example, \cite{paterson2005atmospheric, smith2006two}. However, those studies did not investigate general full-field sets of modes (but mainly the subset of OAM modes) or didn't make concrete predictions about the basis-dependence that only Ince-Gaussian modes allows.

\section{Conclusion}
We employ a state-of-the-art numerical simulation model and use it to investigate the propagation of very general spatial modes of light in realistic conditions. Therewith, we find the interesting effects that the atmosphere is basis-dependent. That means information encoded in one complete basis of spatially-modulated beams can be transmitted with different quality depending on which basis one uses. This is interesting as all complete spatial bases cover the same space, and modes of one basis can be expressed as a linear combination of modes in another basis. What is more, the influence of this effect can be up to 7\% of the total transmission quality, which makes it suitable for experimental observation. To observe this result, we describe an experiment which is feasible with today's technology and communication channels that have already been established in several locations worldwide.

Finally, in order to reduce the gap between theoretical understanding of atmospheric models and its empirical investigations, we would like to advocate a scientific program, which we call \textit{the phenomenology of complex atmospheric light propagation}: Here, the goal is to distill new, empirically observable predictions of light propagation from the best existing atmospheric turbulence models. Those results will allow physicists to perform interesting real-world, outdoor experiments to observe novel, potentially unintuitive phenomena.

In particular phenomena that distinguish between different models, or show the effect of (the absents of) certain theoretical approximations could inspire real-world experimental campaigns. Especially proposals for the strong-light regime could be accessible to state-of-the-art technology.

An interesting question that follows from our study is ``What set of mode is most robust against influence of atmospheric turbulence". The behavior of controlled propagation through scattering material \cite{rotter2017light} and eigenmodes of multi-mode fibers \cite{carpenter2017comparison} are well understood. However, the time-dependence of turbulence (in the order of 1/100sec) makes it a challenging dynamic (and computational expensive) optimization problem.

Likewise, publically sharing of systematically recorded experimental outdoor data, which is not available at all today, could further help the modeling of real-world phenomena on a theoretical level. For the sharing of data, well-established scientific data-sharing platforms such as \textit{datadryad} or scientific data journals such as \textit{Nature: Scientific Data} are adequate resources \cite{candela2015data}.

We believe that this \textit{program} could lead to flourishing experimental endeavours, which -- apart from its pure scientific purpose -- might have an impact in practical questions such as classical and quantum communication, or potentially novel methods to measure properties of atmosphere and thereby weather dynamics.

\section*{Acknowledgments}
The authors thank Mitchell A. Cox, Giovanni Milione, Giacomo Sorelli, Anton Zeilinger and Johannes Handsteiner for useful discussions and valuable comments on the manuscript. The authors thank Dominik Leitner for providing computational resources. XG and LC acknowledge support from National Key Research and Development Program of China (2017YFA0303700); The Major Program of National Natural Science Foundation of China (No. 11690030, 11690032); The National Natural Science Foundation of China (No.61771236); China Scholarship Council Scholarship (CSC). M.K. acknowledges support from the Austrian Science Fund (FWF) through the Erwin Schr\"odinger fellowship No. J4309. This work was supported by the Austrian Academy of Science (\"OAW) and from  Austrian Science Fund (FWF) with SFB F40 (FOQUS).

\clearpage
\appendix
\begin{widetext}
\section{Additional details on Numerical Simulation}
We show some values of the mode cross-talk matrix of IG modes $IG_{4,0,\epsilon}$ in Tables. \ref{tab:CorssTalkMatrixsmall} and \ref{tab:CorssTalkMatrixlarge}. We can see that the mode cross-talk becomes less for radial-like modes by increasing $\epsilon$ and the fidelity of the propagated mode (highlight in green) is significantly larger than that with small $\epsilon$, as described in the main text in Fig. \ref{fig:Possiblereason}.

We show each mode fidelity for HG modes in different order $M$ in Table. \ref{tab:OnlyIndexZeroHG}. We can clearly see that when one of the index of HG mode is zero, the mode performs better than other modes in the same order. The average fidelity for each modes are over 2000 iterations of propagations. The unit for fidelity is percentage and the error is described by standard deviation of the mean. 

In order to verify the mode-dependent effect described in the article is not a numerical artefact, we perform sanity checks involving the analyzation of the normalization, decomposition and orthogonality of different spatial modes in different basis and particularly perform a calculation of pure LG modes comparing to pure HG modes for the same order $M=2$ with a even large grid number N=2048 in Table. \ref{tab:fidelityOrderDif}. There the average fidelity is over all the modes from the same basis in the same order, such as order $M=2$ contains 3 different modes (each modes with more than 3000 iterations). We see those values are within the statistical uncertainty to the values in the main text and show the effect described in the article. Our chosen discrete parameter fulfills the geometrical and aliasing constraints for numerical beam propagation \cite{schmidt2010numerical, coy2005choosing, tang2015propagation}. Therefore we are confident that the results presented in the paper are due to the model instead of the numerical inaccuracy. 

Furthermore, we rotate the HG modes and IG modes with higher ellipticity $\epsilon=4$ by the angle $\phi=45$ degree and propagate these modes through atmospheric turbulence. Then we compare with the LG modes and un-rotated HG or IG modes to further indicate our result is a basis-dependence effect. There we only consider the modes in the same order $M=2$ and the fidelity is over all the modes in the same order from one basis. Due to the expensive computation resource, we only calculate around 1400 samples for rotated HG modes with N=1024 and 1000 samples for rotated IG modes with N=2048 (for reducing the numerical error introduced by rotating matrix) . The results are described in Table. \ref{tab:RotatingHG} and \ref{tab:RotatingIG}. Interestingly, our results confirm that the HG and IG modes with higher ellipticity perform better than the LG modes through turbulence, and these values are within the statistical uncertainty. Our results indicate that rotating these HG and IG modes has no influences to our presented results in the main text, which further predicts the basis-dependence effect is not a numerical artefact but a real result of the model.
 \begin{table}[h]
 	\centering
 	\caption{The average fidelity for mode cross-talk matrix of $IG_{4,0,\epsilon}$ in the case of $\epsilon\rightarrow0$. The fidelity unit is \%.}
 	\begin{tabular}{|c|c|c|c|c|c|}
 		\hline
 		\diagbox[width=1cm]{$n$}{$\ell$}  & -2 & -1 & 0 & 1 & 2  \\ \hline
 		0&0.80$\pm$0.012& 1.31$\pm$0.017& 1.21$\pm$0.018& 1.32$\pm$0.018& 0.77$\pm$0.011\\ \hline
 		1&4.72$\pm$0.052& 4.54$\pm$0.060& 5.13$\pm$0.049& 4.61$\pm$0.063& 4.64$\pm$0.052\\ \hline
 		2&1.44$\pm$0.020& 5.28$\pm$0.073& \cellcolor{mycolor}15.47$\pm$0.224& 5.41$\pm$0.073& 1.46$\pm$0.021\\ \hline
 		3& 0.74$\pm$0.011& 2.13$\pm$0.025& 4.63$\pm$0.049& 2.15$\pm$0.025& 0.76$\pm$0.011\\ \hline
 		4& 0.37$\pm$0.006& 0.85$\pm$0.012& 1.65$\pm$0.020& 0.86$\pm$0.012& 0.36$\pm$0.006\\ \hline
 	\end{tabular}
 	\label{tab:CorssTalkMatrixsmall}
 \end{table}
 \begin{table}[!t]
 	\centering
 	\caption{The average fidelity for mode cross-talk matrix of $IG_{4,0,\epsilon}$ in the case of $\epsilon=4$. The fidelity unit is \%.}
 	\begin{tabular}{|c|c|c|c|c|c|}
 		\hline
 		\diagbox[width=1cm]{$n$}{$\ell$}  & -2 & -1 & 0 & 1 & 2  \\ \hline
 		0& 0.37$\pm$0.007& 1.41$\pm$0.023& 0.96$\pm$0.022& 1.42$\pm$0.024& 0.37$\pm$0.007\\ \hline
 		1& 2.11$\pm$0.029& 5.96$\pm$0.071& 5.53$\pm$0.070& 5.95$\pm$0.072& 2.07$\pm$0.028\\ \hline
 		2&1.74$\pm$0.027& 7.39$\pm$0.088& \cellcolor{mycolor} 21.39$\pm$0.271& 7.60$\pm$0.091& 1.78$\pm$0.027\\ \hline
 		3& 0.68$\pm$0.011& 2.45$\pm$0.031& 5.75$\pm$0.072& 2.46$\pm$0.031& 0.70$\pm$0.011\\ \hline
 		4&0.31$\pm$0.005& 0.96$\pm$0.015& 2.13$\pm$0.033& 0.98$\pm$0.016& 0.31$\pm$0.005\\ \hline
 	\end{tabular}
 	\label{tab:CorssTalkMatrixlarge}
 \end{table}

 \begin{table}[!t]
	\centering
	\caption{The average fidelity for $HG_{n_{x},n_{y}}$ (each mode with 2000 iterations). The fidelity unit is \%. The results shows that HG modes with one index ($n_{x}=0$ or $n_{y}=0$) perform better than other modes in the same order number $M$.}
	\begin{tabular}{|c|c|c|}
		\hline
		order: $M=n_{x}+n_{y}$ &  HG mode: $HG_{n_{x},n_{y}}$& average fidelity $F$ (in \%)\\ \hline
		\multirow{3}{*}{} &$HG_{0,2}$&27.88$\pm$0.50\\ 
		\cline{2-3}
		$M=2$&$HG_{1,1}$&23.87$\pm$0.47\\
		\cline{2-3}
		&$HG_{2,0}$&27.94$\pm$0.50\\  \hline
		\multirow{4}{*}{}&$HG_{0,3}$&24.37$\pm$0.47\\
		\cline{2-3}
		$M=3$ &$HG_{1,2}$&19.81$\pm$0.42\\ 
		\cline{2-3}
		&$HG_{2,1}$&19.65$\pm$0.42\\ 
		\cline{2-3}
		&$HG_{3,0}$&24.37$\pm$0.46\\ \hline
		\multirow{5}{*}{}&$HG_{0,4}$&22.09$\pm$0.44\\
		\cline{2-3}
		&$HG_{1,3}$&17.32$\pm$0.39\\ 
		\cline{2-3}
		$M=4$ &$HG_{2,2}$&16.32$\pm$0.37\\ 
		\cline{2-3}
		&$HG_{3,1}$&17.21$\pm$0.38\\
		\cline{2-3}
		&$HG_{4,0}$&22.00$\pm$0.43\\ \hline     
		\multirow{6}{*}{}&$HG_{0,5}$&20.41$\pm$0.41\\
		\cline{2-3} 
		&$HG_{1,4}$&15.65$\pm$0.36\\ 
		\cline{2-3} 
		&$HG_{2,3}$&14.34$\pm$0.34\\ 
		\cline{2-3}
		$M=5$&$HG_{3,2}$&14.30$\pm$0.34\\
		\cline{2-3}
		&$HG_{4,1}$&15.59$\pm$0.35\\   
		\cline{2-3}
		&$HG_{5,0}$&20.20$\pm$0.40\\ \hline 
	\end{tabular}
	\label{tab:OnlyIndexZeroHG}
\end{table}
\begin{table}[!t]
	\centering
	\caption{Here we perform a very careful test where the size of the sender and receiver plane  are $L1=L_{n}=0.5m$. We calculate all three modes of mode order $M=2$ from LG and HG basis propagating through turbulence with the grid number $N=2048$. The average fidelity is over all the modes in the same order. The results are the same as those in the main text within the statistic uncertainty. In particularly, we also see our main finding -- the basis-dependent effect. This is a clear indication that the effect is not a numerical artefact but a real result of the model.}
	\begin{tabular}{|c|c|c|c|}
		\hline
		\multirow{2}{*}{distance $z$ (m)} & \multicolumn{2}{c|}{average fidelity $F$ (in $\%$)} & \multirow{1}{*}{$\triangle F_{HG-LG}$}\\
		\cline{2-3}
		& LG  & HG & (significance: $\sigma$) \\ \hline
		500& 55.37$\pm$0.24  & 56.93$\pm$0.26   &   1.56  $\pm$0.35 (4.4$\sigma$) \\ \hline
		1000& 34.63$\pm$0.25  & 37.27$\pm$0.26 &   2.64$\pm$0.36 (7.3$\sigma$) \\ \hline
		1500& 23.77$\pm$0.21  & 26.58$\pm$0.22 &   2.81$\pm$0.31 (9.1$\sigma$) \\ \hline
	\end{tabular}
	\label{tab:fidelityOrderDif}
\end{table}
\begin{table}[!t]
	\centering
	\caption{Rotating the HG mode with angle $\phi=45$ degree and compare with LG mode and un-rotated HG modes for order $M=2$. The average fidelity is over all the modes in the same order and fidelity unit is \%. The original data for LG and HG modes are around 6000 samples (N=1024)  while the $HG_{\phi=45}$ is with 1395 sample point under N=1024.}
	\begin{tabular}{|c|c|c|c|c|}
		\hline
		\multicolumn{3}{|c|}{average fidelity $F$ (in $\%$)} & \multirow{1}{*}{$\triangle F_{HG_{\phi=45}-LG}$}&\multirow{1}{*}{$\triangle F_{HG_{\phi=45}-HG}$}\\
		\cline{1-3}
		LG & HG & HG ($\phi=45$)&(significance: $\sigma$) &(significance: $\sigma$)\\ \hline
		23.35$\pm$0.27& 26.57$\pm$0.28& 26.95 $\pm$0.59&3.60$\pm$0.65 (5.54 $\sigma$)&0.38$\pm$0.65 (0.58$\sigma$)\\ \hline
	\end{tabular}
	\label{tab:RotatingHG}
\end{table}
\begin{table}[!t]
	\centering
	\caption{Rotating the IG mode ($\epsilon=4$) with angle $\phi=45$ degree and compare with LG mode and un-rotated IG modes ($\epsilon=4$) for order $M=2$. The average fidelity is over all the modes in the same order and fidelity unit is \%. The original data for LG and IG modes are around 6000 samples (N=1024) while the $IG_{\phi=45}$ is with 1077 sample points under N=2048.}
	\begin{tabular}{|c|c|c|c|c|}
		\hline
		\multicolumn{3}{|c|}{average fidelity $F$ (in $\%$)} & \multirow{1}{*}{$\triangle F_{IG_{\phi=45}-LG}$}&\multirow{1}{*}{$\triangle F_{IG_{\phi=45}-IG}$}\\
		\cline{1-3}
		LG & IG & IG ($\phi=45$)&(significance: $\sigma$) &(significance: $\sigma$)\\ \hline
		23.35$\pm$0.28 & 25.23$\pm$0.28& 25.71$\pm$0.65&2.36$\pm$0.70 (3.37 $\sigma$)&0.48$\pm$0.71 (0.68$\sigma$)\\ \hline
	\end{tabular}
	\label{tab:RotatingIG}
\end{table}
\end{widetext}

\clearpage
\bibliographystyle{unsrt}
\bibliography{refs}

\end{document}